\documentclass[a4paper,fleqn,usenatbib]{mnras}

\usepackage{graphicx}	% Including figure files
\usepackage{amsmath}	% Advanced maths commands
\usepackage{amssymb}	% Extra maths symbols
\usepackage{multicol}        % Multi-column entries in tables
\usepackage{bm}		% Bold maths symbols, including upright Greek
\usepackage{pdflscape}	% Landscape pages
\usepackage{longtable,lscape}
\usepackage{rotating}
\usepackage{multirow,multicol}%%%%%%%%%%%%%%%%%%%%%%%%%%%%%%%%%%%%%%%%%%%%%%%%%%

\newcommand{\kepler}{\textit{Kepler}}
\newcommand{\ktwo}{\textit{K2}}
\newcommand{\ksc}{{\sc k2sc}}
\newcommand{\ksf}{{\sc k2sff}}

\usepackage[T1]{fontenc}
\usepackage{ae,aecompl}

\usepackage{txfonts}
\title[\ktwo\ Campaigns 5 and 6 Transit Candidates]{Transiting exoplanet candidates from \ktwo\ Campaigns 5 and 6}

\author[Pope, Parviainen \& Aigrain]{Benjamin J. S. Pope,\thanks{E-mail: benjamin.pope@physics.ox.ac.uk}, Hannu Parviainen, Suzanne Aigrain
\\
Oxford Astrophysics, University of Oxford, Denys Wilkinson Building, Keble Rd, Oxford OX1 3RH, UK\\
}

\date{Accepted XXX. Received YYY; in original form ZZZ}

\pubyear{2015}

\begin{document}
\label{firstpage}
\pagerange{\pageref{firstpage}--\pageref{lastpage}}
\maketitle

\begin{abstract}
We introduce a new transit search and vetting pipeline for observations from the \ktwo\ mission, and present the candidate transiting planets identified by this pipeline out of the targets in Campaigns~5 and~6. Our pipeline uses the Gaussian Process-based \ksc\ code to correct for the \ktwo\ pointing systematics and simultaneously model stellar variability. The systematics-corrected, variability-detrended light curves are searched for transits with the Box Least Squares method, and a period-dependent detection threshold is used to generate a preliminary candidate list. Two or three individuals vet each candidate manually to produce the final candidate list, using a set of automatically-generated transit fits and assorted diagnostic tests to inform the vetting. We detect 147 single-planet system candidates and 5 multi-planet systems, independently recovering the previously-published hot~Jupiters EPIC~212110888b, WASP-55b (EPIC~212300977b) and Qatar-2b (EPIC~212756297b).
We also report the outcome of reconnaissance spectroscopy carried out for all candidates with \kepler\ magnitude $Kp \le 13$, identifying 12 targets as likely false positives. We compare our results to those of other \ktwo\ transit search pipelines, noting that ours performs particularly well for variable and/or active stars, but that the results are very similar overall. All the light curves and code used in the transit search and vetting process are publicly available, as are the follow-up spectra. 
\end{abstract}


%%%%%%%%%%%%%%%%%%%%%%%%%%%%%%%%%%%%%%%%%%%%%%%%%%

%%%%%%%%%%%%%%%%% BODY OF PAPER %%%%%%%%%%%%%%%%%%

\section{Introduction}
\label{intro}

The \kepler~Space Telescope, operated by NASA, has discovered the majority of known exoplanet candidates, with 4175 planet candidates identified over its 47-month nominal mission \citep{2015ApJS..217...31M}. \kepler\ obtained high-precision ($\sim$~tens of ppm) photometry of $\sim 170,000$ stars over a field of view of 115 $\deg^2$, identifying exoplanets by the characteristic repeated dip in brightness as they transit their host star \citep{2010Sci...327..977B,2010ApJ...713L..79K}.
After the failure of two of the four reaction wheels which formerly stabilized its pointing, it has been revived as the two-wheeled mission \ktwo, using the two remaining wheels with the radiation pressure of the Sun to balance the third axis of rotation. \ktwo's observing sequence consists of a succession of $\sim 70$--80~day `Campaigns', each targeting a field in the Ecliptic plane \citep{2014PASP..126..398H}, with photometric precision comparable to that of the nominal \kepler~mission \citep{2014PASP..126..948V}. As a result of the short Campaigns, \ktwo\ is unable to detect planets with periods as long as were accessible with \kepler. On the other hand, by covering many different fields it is able to search for short-period planets around a larger number of bright stars and to probe a wider range of environments.

\ktwo\ has diverse science goals (see \citealt{2014PASP..126..398H} for a more detailed discussion). It vastly increases upon the number of bright Sun-like stars and nearby late-type stars which were included in the \kepler\ prime mission. Short-period transiting planets discovered around these stars will be particularly good targets for future atmosphere studies. \ktwo\ is also advancing stellar astrophysics through the study of pulsating stars, eclipsing binaries and transients, as well as observing extragalactic and Solar System targets. By surveying fields located in or close to the Ecliptic plane, \ktwo\ observations cover several nearby, young open clusters and associations, such as $\rho$~Ophiuchi ($\rho$~Oph) and Upper Scorpius (Upper Sco) (Campaign~2), the Pleiades (Campaign~4), Praesepe (or the Beehive Cluster) (Campaign~5), and Taurus and the Hyades (Campaigns~4~\&~13). \ktwo\ observations of these clusters are opening a new window on young star variability, including accretion-related variability, stellar activity and rotation, and pulsations, but enable also the discovery and characterisation of young eclipsing binaries and transiting planets. The latter are particularly important probes of the early stages of planetary evolution. 

In this paper we present a new pipeline to correct \ktwo\ light curves for instrumental systematics and stellar variability, search them for planetary transit candidates, and perform a series of diagnostic tests to weed out false posisitives. We apply this pipeline to Campaigns~5 \& 6 (hereafter C5 and~C6), identifying 77 and 71 single-planet candidates among the 25139 and 28291 targets surveyed in each Campaign, as well as 5 systems showing multiple sets of transits. While our method is applicable to any \ktwo\ light curve, and indeed readily adaptable to data from other instruments, a particular focus of its design has been to ensure good performance for young, variable stars, which represent a larger fraction of the most interesting targets for \ktwo\ than for previous transit search missions.

\ktwo\ has already led to a number of noteworthy planetary transit discoveries, including: disintegrating rocky planetesimals transiting a white dwarf \citep{2015Natur.526..546V} and an M-dwarf \citep{2015ApJ...812..112S}, two unexpected additional planets in the hot Jupiter-hosting system WASP-47 \citep{2015ApJ...812L..18B}, planetary companions to field M-dwarfs \citep{2015ApJ...804...10C,2016ApJ...818...87S,2016ApJ...820...41H}, transiting planets orbiting young stars in the Hyades and Upper~Sco \citep{2016ApJ...818...46M,2016arXiv160406165M} and many more individual planets. Several groups have also published catalogues of \ktwo\ planet candidates, including \citet{2015ApJ...806..215F} (Campaign~1), \citet{2016ApJS..222...14V} (Campaigns~0 to~3), \citet{crossfieldcatalog} (Campaigns~0 to~4), \citet{2016arXiv160306488A} (Campaigns~0 to~5). In addition to this, \citet{2015IAUGA..2258352F} and \citet{2016MNRAS.457.2273O} have presented systematic searches for single-transit events in K2, identifying  long-period planet candidates. Where there is overlap, we will compare our results to these published catalogs. 

The different pipelines used on \ktwo\ data to date differ in subtle ways from each other, and from ours, at every stage of the light curve detrending, transit search and vetting process, but particularly important differences are found in the light curve extraction and detrending step. This is because of the  importance of instrumental systematics (and, to a lesser extent, stellar variability) in \ktwo\ data. In the \ktwo\ mission, the \kepler\ satellite is balanced at an unstable equilibrium, using two reaction wheels and with solar radiation pressure keeping the third axis approximately steady. The third axis orientation must be maintained, however, by small thruster firings every $\sim 6$ hours, moving a typical star by of order  $\sim$~1~pixel across the detector, with the result that \ktwo\ data contain significant pointing-related systematic photometric trends that make robust inference about the presence of planets more difficult than in the nominal \kepler\ mission. Several data-reduction pipelines have been developed by different groups to compensate for this effect \citep{2014PASP..126..948V,2015MNRAS.447.2880A,2015A&A...579A..19A,2015ApJ...806..215F,2015ApJ...806...30L}, including our detrending pipeline (which we use in this paper) \citep[][hereafter APP16]{k2sc}. A unique feature of our approach is that systematics and stellar variability are modelled simultaneously rather than sequentially, which significantly improves the results when both effects are significant, especially on similar timescales. The methods used in the remainder of our pipeline for transit search and vetting are relatively standard, but were implemented with particular attention to scalability and computational efficiency, and in such a way as to enable easy, uniform human vetting at the present time, as well as -- in the future -- progression towards a fully probabilistic asssessment of the likelihood that individual candidates are \emph{bona fide} planets.

The remainder of this paper is structured as follows. Sections~\ref{syscor}, \ref{search} and \ref{vetting} describe our light curve preparation, transit search and vetting methodology, respectively. Our catalog of transiting planet candidates from C5 and C6 is given in Section~\ref{candidates}, together with the results of preliminary spectroscopic reconnaissance. Section~\ref{code} gives links to the code used in our pipeline, all of which is publicly available, and we give our conclusions and discuss plans for future improvements in Section~\ref{conclusions}.

\section{Light Curve Preparation}
\label{syscor}

\ktwo\ photometric time series are typically affected by sawtooth-like variations in measured flux, due to the drift and subsequent thruster reset of the spacecraft boresight roll angle. This effect, needed to maintain spacecraft pointing near the unstable equilibrium induced by solar radiation pressure, causes each star to move by of order $\sim 1$ pixel over the course of each thruster-reset period of $\sim 6$~hours. This is translated into variations in flux by the differential sensitivity between pixels (inter-pixel variations) and across the surface of each pixel (intra-pixel variations), and by the loss of light outside the photometric aperture as it imperfectly matches and tracks the point spread function (aperture losses). Other systematic effects are also present (as they were in the light curves from the original \kepler\ mission), but the pointing-related variations are by far the most prominent. In addition to these instrumental systematics, the light curves also contain intrinsic stellar variability, most commonly due to the rotational modulation of star spots. Both systematics and variability can severely hinder the detection of planetary transits, and must be filtered or modelled before the transit search can proceed. 

The standard approach until now has been to do this sequentially, first modelling the systematics while ignoring the variability, then using some variation on a high-pass filter to remove most of the variability while preserving any planetary transits (the latter step exploits the fact that transits occur on shorter timescales than most forms of stellar variability). However, this approach breaks down when both systematics and variability are of comparable amplitude, and occur on similar timescales. While the typical rotation periods of \ktwo\ target stars (days to weeks) are significantly longer than the $\sim 6$ hour characteristic timescale of the pointing variations, there is nonetheless significant power in the variability of many stars on 6 to 12-hour timescales. This is particularly problematic for \ktwo, which observes a larger fraction of young, rapidly rotating, active stars than earlier transit-search missions. In such cases, modelling the systematics and variability simultaneously tends to be more effective, particularly if -- as in the case of \ktwo\ -- the systematics can be constrained to depend primarily on the star's position on the detector, which should have no impact on its intrinsic brightness. This is the approach implemented in the \ksc\ pipeline APP16, which uses a Gaussian process to model the systematics and variability simultaneously. The model contains two distinct components: the first depends on position only and represents the pointing-related systematics, and the second depends on time only, and represents the variability (plus any residual systematics not related to position). A sophisticated outlier rejection scheme is used to ensure that flares, eclipses and transits do not adversely affect the modelling, but are nonetheless preserved. We model the stellar variability one of two ways, either with the quasi-periodic exponential sine squared kernel in time  \citep{Rasmussen:2005:GPM:1162254} or with a squared exponential, and decide between the two by calculating a Lomb-Scargle periodogram \citep{1976Ap&SS..39..447L} of the input lightcurve; if this shows a strong periodicity, we choose the quasi-periodic kernel. Otherwise, we represent stellar variability with a slowly-varying squared exponential kernel.

We begin with the \ktwo\ light curves available at the Mikulski Archive for Space Telescopes (MAST), which we process with \ksc\ to remove both systematics and variability. We exclude EPICs 200008644~\textendash~200009280 in C5, which correspond to cluster superstamps (large contiguous regions of active pixels that cover star clusters) and to trans-Neptunian objects, and 200041889~\textendash~200061149 in C6, which track Trojan asteroids. While cluster science is a key science goal of \ktwo, standard MAST-pipeline aperture photometry is not reliable in the extremely crowded fields of the cores of clusters, as are treated with superstamps. Many more widely distributed cluster members and probable cluster members outside the superstamp have MAST lightcurves; we exclude from the analysis presented here stars with very high cluster membership probability, in particular Praesepe stars (C5), as the subject of a separate study. 

In APP16, we tested \ksc\ on both the Simple Aperture Photometry (SAP) light curves, and light curves corrected for common-mode systematic trends  by the \kepler\ Presearch Data Conditioning (PDC) pipeline \citep{2010SPIE.7740E..1UT,2012PASP..124..985S,2012PASP..124.1000S}. We found that the latter typically result in slightly improved photometric precision on transit timescales (as measured using our proxy estimate of the 6.5-h combined differential photometric precision, or CDPP). This is most likely due to the fact that some of the common-mode trends removed by the PDC pipeline are related to variations in (e.g.) telescope focus and the temperature of various components, and not directly to the pointing, and are thus not modelled by \ksc. On the other hand, \ksc\ significantly improves the precision of PDC light curves, and enhances our ability to detect planetary transits (as demonstrated by APP16 using transit injection tests). We therefore use the \ksc-processed PDC light curves as our starting point for the transit search. The interested reader is referred to APP16 for more details on \ksc\ and its application to \ktwo\ light curves.

\begin{figure*}
\centering
\includegraphics[width=\linewidth]{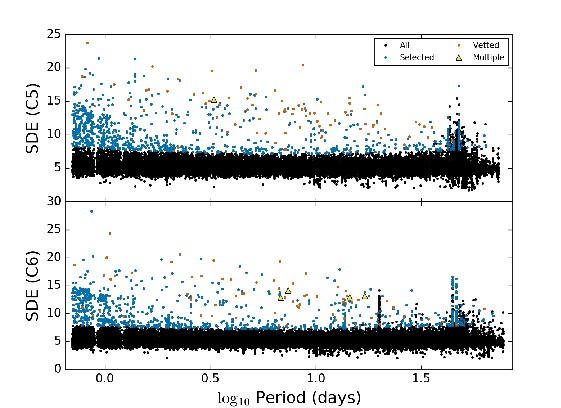}
\caption{Initial selection of transit candidates based on the BLS detection statistic and period. We display SDE versus period for all objects in black, for C5 (top) and C6 (bottom). EPICs selected for visual examination (as described in Section~\ref{search}) are shown in blue, and targets which pass the vetting process (described in Section~\ref{vetting}) in orange. Yellow triangles display the first detected planet for the multiple-planet systems described in Section~\ref{multiplanets}.  
}
\label{candidates_combined}
\end{figure*}

\section{Transit Search}
\label{search}

Candidates are initially identified by a Box-Least-Squares (BLS) algorithm \citep{2002A&A...391..369K} which evaluates a signal-to-noise statistic, the signal detection efficiency (SDE), for a simple box-shaped transit for a grid of zero epochs spanning the Campaign and for periods ranging from 0.7~d to 98\% of the total duration of the Campaign, using our own version of the algorithm \citep{Parviainen2016BLS}. We apply this search to a GP-detrended light curve, which has been prewhitened, where we subtract the GP time component to model smooth out of transit variations \citep{2003A&A...401..743C}. We record the SDE, and best-fitting period, epoch and duration for each target.

Using this preliminary fit, we then fit a transit model using the \textsc{PyTransit} package \citep{Parviainen2015pt}, a fast implementation of the \citet{2002ApJ...580L.171M} analytic transit light curves.

We use the BLS SDE to identify promising candidates for more detailed study. Using the injection test results from APP16~(Figure~10), we expect non-planet-hosts to be distributed around a mean SDE of $\sim 6$, roughly uniformly in log-period. In APP16 we therefore recommended a cut in SDE of $\sim 8$ as a simple and effective way of selecting transit candidates. As is apparent in Fig.~\ref{candidates_combined}, in our real data, there is in fact some slope to this distribution, as well as a larger number of candidates at short periods. 

In order to deal with this, we define an SDE threshold as a function of period. First, we discard single transit candidates as beyond the scope of this study, and examine the remaining objects. In preliminary inspection, we note that $\sim 65$ long-period candidates in C6 result from systematics associated with particular pairs of bad epochs, leading to a pileup of candidates with poorly-corrected lightcurves with zero epochs in two $\sim 0.5$~d windows. We therefore identify these from a histogram of fitted BLS zero epochs, and remove these candidates before proceeding. We also discard candidates for which the log-likelihood of a sinusoidal fit is greater than that of a transit, as a way of removing the most obvious variable sources. We then split the objects into 50 bins in period space, each containing an equal number of objects (469 for C5 and 457 for C6), thus using a variable bin size. In each bin we find the median SDE, and the median absolute deviation-estimated standard deviation. Since there is a trend in the median as a function of period, we fit a second-order polynomial to the distribution of period as a function of median SDE, subtract this from each, and find the standard deviation of the residuals. We then select for further analysis all candidates that lie more than $3~\sigma$ above this value in each bin. We illustrate this in Figure~\ref{candidates_combined}, displaying a plot of SDE versus period for both Campaigns, including all objects, selected objects, and vetted objects.

\section{Vetting}
\label{vetting}

The procedure described in the previous section leads to a preliminary candidate list for each Campaign, which still contains both false detections caused by light curve artefacts or other abnormalities, and astrophysical false positives, i.e. transit-like events which are not caused by a planet. To identify these and produce a cleaner candidate list, we visually inspect the light curves, along with a set of diagnostic plots based on transit model fits, broadly following the approach outlined in \citet{2010ApJ...713L.103B}. 

\begin{figure*}
\centering
\includegraphics[height=0.9\textheight]{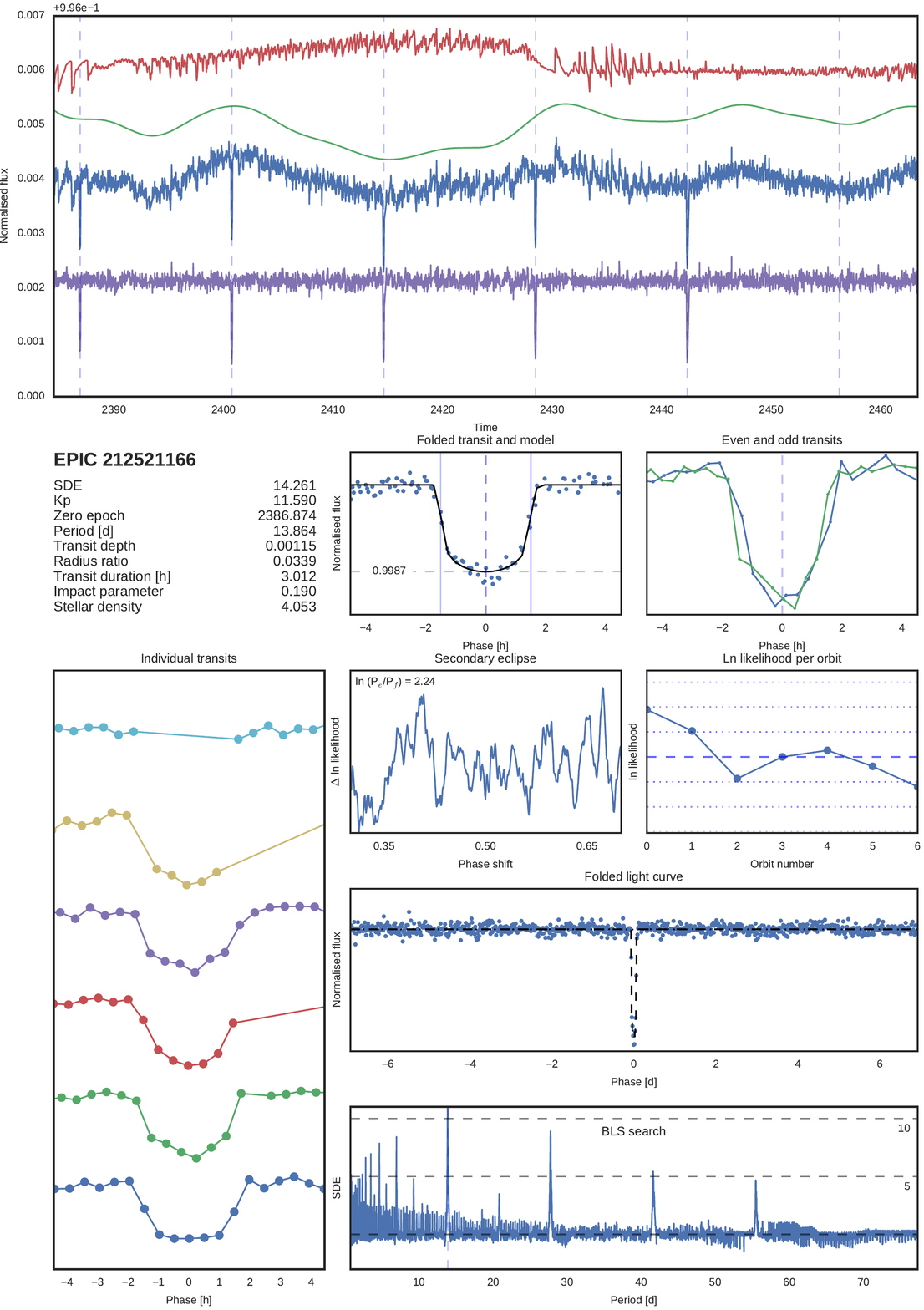}
\caption{Example of a single-page vetting report for one of our candidates. See text for details of the individual panels.}
\label{vetting_fig}
\end{figure*}

Some light curves are affected by poorly-corrected systematics. These show residual $\sim 6$ hour variability which can lead to a spuriously large BLS detection statistic. We therefore systematically discard detections at the characteristic thruster firing period, and multiples thereof. Furthermore, the time component of the \ksc\ GP model does not always model very short-period or high-amplitude stellar variability well. This is especially true for classical pulsators such as RR~Lyr and Cepheid variables: while the overall variability pattern is modelled adequately, the extrema are not, leaving periodic dips in the residuals, which can lead to a high BLS statistic. Finally, while planetary transits and short-duration stellar eclipses are typically flagged as outliers by the \ksc\ model (so that they are left unchanged by the detrending process), the outlier flagging procedure is less successful for high-amplitude eclipsing binaries (EBs) where the eclipse duration exceeds a few percent of the period. In those cases, the time component of the GP model partially reproduces the eclipses, leaving in the residuals shallow, periodic eclipse-like events which can superficially resemble planetary transits. All of these categories of false positive are readily identified by visually comparing the raw light curve, the systematics and time-dependent components of the \ksc\ GP model, and the detrended light curve. 

Eclipsing binaries (EBs) are the major source of false positives remaining after this step. The different types of EBs that can mimic a transit signal are discussed extensively in the literature (see e.g.\ \citealt{2004AIPC..713..151C} for an overview); we merely list them here: grazing EBs, EBs containing a large primary or small secondary star, and blended EBs (EBs whose eclipses are diluted by the light of a third star enclosed in the photometry aperture). To aid in the identification of such systems, we systematically produce, for all our candidates, a single-page report displaying key diagnostics in graphical and numerical form, as listed below. An example of such a report is shown in Figure~\ref{vetting_fig}, the reports for all the candidates listed in Section~\ref{candidates} are provided online as supplementary material. The report shows, from top left to bottom right:
\begin{itemize}
\item\emph{Unfolded light curve plots} showing the raw light curve, the systematics and time-component of the \ksc\ GP model, and the detrended light curve used in the BLS transit search. The location of the transits is indicated by vertical dashed lines (if the light curve contains $>20$ transits, only the first 20 are marked);
\item \emph{Main BLS and transit fit parameters} giving the BLS SDE detection statistic, the $Kp$ magnitude of the target (magnitude in the \kepler\ bandpass), and the best-fitting parameters of a transit fit to the phase-folded detrended light curve;
\item \emph{Folded transit plot} showing the best-fit model overlaid on the phase-folded light curve;
\item \emph{Even and odd transit plots} showing, in different colours, the binned phase-folded light curves for even- and odd-numbered transits (helpful in identifying near-equal mass EBs where the BLS-detected period is half the true orbital period);
\item \emph{Individual transit plots} showing each transit (or the first 20 if the total number exceeds 20) separately (helpful in identifying false detections caused by light curve artefacts);
\item \emph{Secondary eclipse detection plots} showing the log-likelihood of a secondary eclipse with the same parameters as the best-fit transit model, but variable depth, as a function of phase from 0.3 to 0.7 (EBs with low to moderate eccentricities but detectable secondary eclipses cause a clear peak in such a diagram);
\item \emph{Per-orbit transit likelihood plot} showing the log-likelihood of the best-fit transit model for each transit event separately (strong variations from one transit to the next indicate that the detection may be caused by or affected by light curve artefacts);
\item \emph{Full folded light curve plot}: this can reveal secondary eclipses occurring outside the 0.3--0.7 phase range, or additional transit events
\item \emph{BLS periodogram} providing an additional means of assessing the significance of the detection (subsidiary peaks can also reveal additional sets of transits in the light curve).
\end{itemize}

All the transit model fits are performed using the \textsc{PyTransit} package, which implements the analytic model of \citet{2002ApJ...580L.171M}. The parameters of the transit fits are epoch, period, planet-to-star radius ratio, impact parameter and stellar density. We also report the transit depth and duration, which are computed from the best-fit transit model. Note that the transit models are evaluated with 3\,min time-sampling and binned up by a factor of 10 to account for the smearing of the ingress and egress caused by the relatively sparse sampling of \ktwo\ long cadence observations. 
The vetting process for transit candidates identified during the \kepler\ prime mission (see e.g.\ \citealt{2010ApJ...713L.103B}) also relies on the use of so-called "rain diagrams": plots of flux versus x- and y-position of the target centroid. Such diagrams are highly effective at identifying blended EBs, where the target centroid can change significantly during eclipses (except for hierarchical triple system where the EB and contaminant are essentially co-located). In the case of \ktwo, one must use the difference between the star's predicted position (based on the overall pointing variations of the satellite) and its measured centroid. We systematically produce such plots and use them in the vetting process, but they have not yet been incorporated into the single-page reports.

Any candidates which show clear signs of being false detections, or being caused by EBs, on the basis of the diagnostics described above, are discarded from the final candidate lists. Rather than define quantitative criteria for rejecting or keeping a candidate, at least two individuals vet all the candidates independently, and only those considered passable by all the vetters are retained. This clearly leaves scope for considerable improvement: ideally we would like to perform a systematic, quantitative assessment of the probability that each candidate is indeed a planet, based on all the available information. One way to do this automatically is to use machine learning tools, such as the random forest algorithm recently implemented for \kepler\ \citep{2015ApJ...806....6M}, trained on transit candidate lists vetted by humans. However, this is beyond the scope of the present paper.

\section{Planet Candidates}
\label{candidates}

\begin{figure}
\centering
\includegraphics[width=0.85\linewidth]{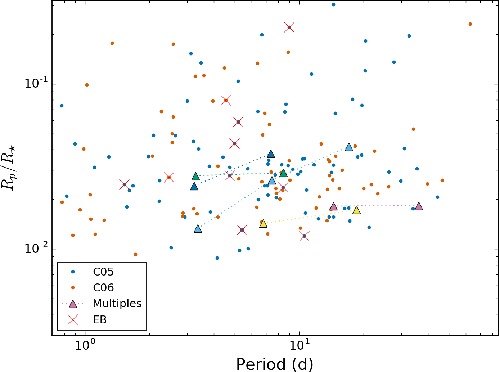}
\caption{Radius ratio versus period for all vetted planet candidates from C5 (blue) and C6 (orange). Multiple systems are shown as triangles, with planets in each individual system sharing the same colour and joined by dotted lines. Objects which were found to be binaries following reconnaissance spectroscopy (see Section~\ref{tres}) are displayed with red crosses superimposed.
}
\label{allplanets}
\end{figure}

In Tables~\ref{table_5} and~\ref{table_6} we list the planet candidates that pass the tests described in Section~\ref{vetting}, for C5 and C6, respectively. The last two columns give the number of reconnaissance spectra obtained for each candidate, where applicable, and the status of the candidate based on the spectra plus any other available information (see Section~\ref{tres} for more details). The period and radius ratio of these candidates is displayed in Figure~\ref{allplanets}. Out of 996 and 981 objects pre-selected for visual inspection in C5 and C6 respectively, we select 87 \& 77 as \emph{bona fide} candidates after vetting. A vetting diagnostic diagram similar to Figure~\ref{vetting_fig} is provided for each candidate in the supplementary online material.

Our C5 candidate list includes the hot Jupiter EPIC~212110888b, which has been independently discovered (from its \ktwo\ light curve) and spectroscopically confirmed by \citet{2016arXiv160107635L} and~\citet{2016arXiv160200638H}. With an $0.8\%$ transit around an F9 star, the two publications agree on a mass of $M_p \sim 1.7 \pm 0.1 M_J$ and a radius of $1.4 \pm 0.1 R_J$, differing within uncertainty according to the different stellar models. In addition to this, \citet{2016arXiv160504291O} have obtained RV spectroscopy of EPIC212521166~b, first reported by \citet{aigrain_2016_45873}, finding it to be a massive mini-Neptune with similar photometric parameters to those reported here.

Several long-period candidates were found around stars whose light curves display stochastic oscillations on timescales of a few hours, and are hence likely to be red giants (EPIC~212411479, 212438212, 211996053, and 212481820). Although they did not explicitly fail any of the vetting tests, we did not include them in the final candidate lists, as it is difficult to distinguish a real transit from residuals from imperfectly modelled, stochastic stellar variability on the same timescale. We note however that these events are in principle detectable, and might warrant more detailed study.

Our target sample partly overlaps with that of \citet{2016arXiv160306488A}, who searched for ultra-short-period planets (up to 1~day only) in Campaigns 0--5. They report list four candidates in C5: of these, we do not pre-select EPIC~211357309 or~211995325 for followup based on the BLS~SDE, while we do preselect 211685045 and~212150006. Both of the planets we miss have periods shorter than the 0.7~d cutoff we impose on our initial search. We did not include shorter periods in our search because both very short period pulsating stars and residual pointing systematics at $4 c/d$ cause many false positives in that regime. EPIC~212150006 is listed in Table~\ref{table_5} as a \emph{bona fide} planet candidate. Our light curve for 211685045 appears to show residual variability consistent with a false positive, so it was rejected at the vetting stage.

We have also compared our list candidate list for both Campaigns to those kindly provided by A.~Vanderburg (priv.\ comm.). As we were co-ordinating our spectroscopic reconnaissance efforts, we shared our targets lists for magnitudes $K_p < 13$. Within that range, there is a very substantial overlap between our lists, with only a small number of objects found by one group but not by the other. In several cases, we have rejected candidates found by Vanderburg as having evidence for a secondary eclipse; in other cases, Vanderburg finds evidence of a secondary eclipse where we do not (see Comments in Tables~\ref{table_5} and~\ref{table_6}). A few of our candidates were not detected by Vanderburg, these correspond to active stars whose variability and systematics are better-modelled by the \ksc\ GP model than by piecewise polynomial detrending. Finally, we failed to detect a few of Vanderburg's candidates. In those cases, we see no evidence for a transit in our light curve, but the signal is clear in the \ksf\ light curves. These are all bright stars, for which we suspect that the \ksf\ photometric apertures are better suited to the point spread function than those used by MAST \citep{2014PASP..126..948V}. 

This comparison shows that the different light curve detrending, transit search and vetting processes are complementary, and that a more complete candidate list may be obtained by combining the results of several pipelines than by any single pipeline. We also note that the overall transit detection potential of \ktwo\ may be best exploited by combining \ksf\ aperture photometry (or similar) with \ksc\ detrending.

\subsection{Multiple Planet Systems}
\label{multiplanets}

We also checked for additional transiting planets in the systems identified in Section~\ref{vetting}. At the vetting stage, we made note of systems for which there was visual evidence of more than one set of transits. For each such system, we then subtracted the best-fit model for the first set of transits and repeated the transit search and fitting on the residuals. The process was iterated until no further sets of transits were found.

We detect two sets of transits around each of EPIC~212012119 (C5); and 212393193, 212703473, and~212779596 (C6), and three sets of transits around 212768333 (C6). The parameters of these systems are presented in Table~\ref{multi_table}. For each of these systems, we provide a vetting diagnostic diagram similar to Figure~\ref{vetting_fig} for each set of transits separately in the supplementary online material.  

EPIC~212651213 and~212651234 show two identical sets of transits which pass our initial vetting procedure, indicating a photometric blend. EPIC~212651213 was later found to be triple-lined (see Section~\ref{tres}), while~212651234 shows no RV variation over 3 epochs. We therefore conclude that EPIC~212651213 is an eclipsing SB3 system, and the apparent transits on 212651234 are the result of a photometric blend. These two systems are therefore not included in Table~\ref{multi_table}. 

In addition to these, we detect three transit-like events with $\sim 3$~to~$5\%$ depth around EPIC~212656205, which do not have a periodic relation and have different depths. These may represent a planar hierarchical multiple stellar system, or multiple close-in giant planets. 

\subsection{Spectroscopic Reconnaissance}
\label{tres}

For most of the objects in the EPIC catalogue, spectra have not previously been obtained and only photometric information is available. A single, moderate to high resolution spectrum can be used to determine the host star's spectral type and evolutionary stage, and to check for obvious signs of binarity (double- or triple-lined spectra). For single-lined, main-sequence cool stars, additional spectra can reveal km/s radial velocity variations, indicative of a stellar or substellar companion. All our candidates which passed the vetting stage and were bright enough for efficient spectroscopic follow-up ($Kp \le 13$) were observed at least once with the Tillinghast Reflector Echelle Spectrograph (TRES), at the 1.5-meter Tillinghast Telescope on Mt Hopkins, Arizona. TRES has three spectroscopic resolving power settings available: 41,000 (high), 30,000 (medium), and 20,000 (low); reconnaissance is typically conducted in the medium resolution mode. In most cases, the radial velocity (RV) precision achievable with TRES ($\sim 0.1 \text{km/s}$ or somewhat better) is not sufficient to detect the signal of a planetary-mass companion to a main sequence star, except for hot Jupiters under favourable conditions.

Where possible, the systems were observed at or near quadrature in order to maximise the chances of resolving multiple sets of stellar lines. Any systems identified as giants, double- or triple-lined binaries based on the first observation were not observed any further. These are noted as "GIANT", "SB2" or "SB3" in the final column of Tables~\ref{table_6} and~\ref{table_6}. Additional spectra were taken near the opposite quadrature for some of the remaining single-lined candidates, if the planet-to-star radius ratio indicated that an orbital solution might be feasible with TRES. If the resulting radial velocity change was too large for a planetary companion, the system was classified as a single-lined stellar binary (labelled "SB1" in the tables). In the case of EPIC~212110888, which contains a hot Jupiter, further additional spectra were obtained in order to obtain a full orbital solution, which is consistent with the parameters published by \citet{2016arXiv160107635L} and~\citet{2016arXiv160200638H}, with TRES spectra best fit by a $166 \pm 21$ RV semi-amplitude planet, with a mass of $1.18^{+0.15}_{-0.15} M_J$. One system (EPIC~212808289) may host a warm Jupiter, but additional spectra, ideally with better radial velocity precision, are needed to confirm this. The inner planet in EPIC~212703473 is also a likely hot Jupiter. For EPIC~212066407 and 212803289 we list RV semi-amplitude $K$ as determined by TRES, while for 211733267 and 211818569, with 2 epochs showing little variation we list upper limits on $K$. The remaining systems included in the TRES reconaissance spectroscopy program which were not found to show any evidence of binarity are simply noted as "OK" in Tables~\ref{table_6} and~\ref{table_6}. For the brightest among them, a full orbital solution might be feasible with dedicated high-precision radial velocity instrument, otherwise a statistical evaluation of the likelihood that the companion(s) is/are indeed planetary ("validation") could be performed based on all the information available to date plus high spatial resolution imaging.

Overall, the TRES reconaissance program enabled us to rule out 8 out of the 86 single-planet candidates from C5, and 2 out of the 71 from C6, as giants or spectroscopic binaries, for a final total of surviving 147 single-planet candidates across both Campaigns. 

\section{Source Code and Data Products}
\label{code}

In the interests of open science, we have our code implementing the methods described in Sections~\ref{syscor}, \ref{search} and~\ref{vetting} publicly available. It consists of several open source packages distributed under a GPL license. We invite interested readers to use, modify and contribute to this code as an evolving resource for the astronomical community. The \ksc\ systematics correction code \citep{k2sc} is available on GitHub at
\begin{itemize}
    \item[] \url{https://github.com/OxES/k2sc},
\end{itemize}
the transit modelling code \textsc{PyTransit} \citep{2015MNRAS.450.3233P} at
\begin{itemize}
    \item[] \url{https://github.com/hpparvi/PyTransit},
\end{itemize}
and the code used to carry out the BLS transit search and produce the candidate reports, \textsc{k2ps}, is available at
\begin{itemize}
    \item[] \url{https://github.com/hpparvi/k2ps}.
\end{itemize}
The data used and produced in the course of this paper are also publicly available. The full set of \ksc-processed \ktwo\ lightcurves is available at MAST
\begin{itemize}
    \item[] \url{https://archive.stsci.edu/prepds/k2sc/},
\end{itemize}
and the candidate reports used in the vetting process are provided with this paper (for surviving candidates only) as supplementary online material, and the TRES spectra of our candidates have been uploaded to the `ExoFOP--K2' website
\begin{itemize}
    \item[] \url{https://exofop.ipac.caltech.edu/k2/}.
\end{itemize}

\section{Conclusions}
\label{conclusions}

We have presented a new pipeline to search for transiting planet candidates in \ktwo\ data, and reported the results of applying this pipeline to  Campaigns~5 \&~6, together with reconnaissance spectroscopy of the brightest of our candidates.  We recover known planets and identify false positives due to stellar-mass companions. We have made our code, light curves, and diagnostic data products publicly available to facilitate confirmation and comparison, and look forward to comparing our results with future work using other methods.

We have compared our results to those by \citet{2016arXiv160306488A}, the only published catalog which covers one of the Campaigns we have analysed (C5), as well as with the unpublished candidate list produced by A.~Vanderburg for both C5 and~C6. In both cases, our results are generally consistent. The two candidates identified by \citet{2016arXiv160306488A} and not by ourselves lie outside our chosen period search range. In several cases, we identified candidates found by Vanderburg, but discarded them at the vetting stage. On the other hand, we also failed to detect a few of his candidates altogether, and these correspond to cases where the photometric precision of our light curves is not as good as that of the \ksf\ light curves used by Vanderburg for the same object.

We also identify a small number of candidates, which were not reported by other teams. These are predominantly orbiting variable stars, which is consistent with our stated intention to write a pipeline that is particularly robust to astrophysical variability. More detailed comparison to other methods will become possible once more groups have published candidate lists for C5 and C6, and once we have processed earlier Campaigns, for which published candidate lists are already available. 

Our results are broadly compatible with the K2SC injection tests presented in APP16, but a direct comparison is not feasible, as we have made some small but significant changes to our transit search methodology since. We will repeat the injection tests at a later date,  after processing more Campaigns, to provide a quantitative assessment of the sensitivity of our final pipeline. 
There are a number of possible modifications to our pipeline that may yield improvements in sensitivity or reliability in the future. First, as noted above, for some bright stars our light curves are not as precise as those used by some other teams. We use the {\sc PDC}-MAP light curves as our starting point; these are extracted using fixed, pixelized photometric aperture masks. For some types of stars (particularly bright stars, as noted in \citealt{2014PASP..126..948V} and \citealt{2015ApJ...806...30L}), these apertures may not be optimal. Better sensitivity to shallow transits might be achieved by applying \ksc\ to light curves extracted using more optimized apertures. Another area where we hope to make progress in the near future is in making the vetting process more automatic, perhaps by implementing and training machine learning algorithms to distinguish between planetary transits and false positives.

Beyond the discovery of individual interesting systems, a significant element of the long-term legacy of the \ktwo\ mission will be improved estimates of short-period planet incidence rates around types of stars which were relatively under-represented in the \kepler\ prime mission (such as M-dwarfs), as well as the ability to check for if the incidence of short-period planets depends in a measurable way on direction within the Ecliptic plane. By making all our code publicly available, we hope to facilitate the evaluation of such incidence rates, by providing all the tools needed for complete end-to-end simulations of the detection process.

\section*{Acknowledgements}

We would like to thank Andrew Vanderburg for his correspondence in discussing our pipeline, comparing results and sharing information on possible false positive candidates. We would also like to thank Dave Latham for his helpful comments and his collaboration in TRES followup of our planet candidates. We are grateful for the efforts of the \kepler\ and~\ktwo\ teams, and in particular to Geert Barentsen for his help in answering some very detailed queries.

This research made use of NASA's Astrophysics Data System; the SIMBAD database, operated at CDS, Strasbourg, France; and the IPython \citep{PER-GRA:2007} and SciPy \citep{jones_scipy_2001} packages. Some of the data presented in this paper were obtained from the Mikulski Archive for Space Telescopes (MAST). STScI is operated by the Association of Universities for Research in Astronomy, Inc., under NASA contract NAS5-26555. Support for MAST for non-HST data is provided by the NASA Office of Space Science via grant NNX13AC07G and by other grants and contracts. 

Financial support for this work came from Balliol College, the Clarendon Fund, the Leverhulme Trust and the UK Science and Technology Facilities Council (STFC).

%%%%%%%%%%%%%%%%%%%%%%%%%%%%%%%%%%%%%%%%%%%%%%%%%%

%%%%%%%%%%%%%%%%%%%%%%%%%%%%%%%%%%%%%%%%%%%%%%%%%%



\bibliographystyle{mnras}
\bibliography{ms} % if your bibtex file is called example.bib

% \newpage

\appendix

% \section{Supplementary Material}

\onecolumn

\begin{center}
\renewcommand{\thefootnote}{\fnsymbol{footnote}}
\renewcommand{\arraystretch}{0.79}
\scriptsize
\begin{longtable}{rrrrrrrrrrrrrr}
\caption[Planet Candidates]{Full list of our 86 vetted single planet candidates in \ktwo\ Campaign~5. RV semi-amplitude $K$ (or limits) listed in Comments where applicable.}
\label{table_5} \\

\hline \hline \\[-2ex]
  \multicolumn{1}{c}{\textbf{EPIC}} &
  \multicolumn{1}{c}{\textbf{RA}} &
  \multicolumn{1}{c}{\textbf{Dec}} &
  \multicolumn{1}{c}{\textbf{V}} &
  \multicolumn{1}{c}{\textbf{J-K}} &
  \multicolumn{1}{c}{\textbf{Period}} &
  \multicolumn{1}{c}{\textbf{Epoch}} &
  \multicolumn{1}{c}{\textbf{Depth}} &
  \multicolumn{1}{c}{\textbf{$r_p/r_\star$}} &
  \multicolumn{1}{c}{\textbf{Duration}} &
  \multicolumn{1}{c}{\textbf{Impact}} &
  \multicolumn{1}{c}{\textbf{Density}} &
  \multicolumn{1}{c}{\textbf{TRES}} &
  \multicolumn{1}{c}{\textbf{Comments}}  \\[0.8ex] 
  \multicolumn{1}{c}{\textbf{  }} &
  \multicolumn{1}{c}{\textbf{  }} &
  \multicolumn{1}{c}{\textbf{  }} &
  \multicolumn{1}{c}{\textbf{(mag)}} &
  \multicolumn{1}{c}{\textbf{(mag)}} &
  \multicolumn{1}{c}{\textbf{(days)}} &
  \multicolumn{1}{c}{\textbf{(BJD)}} &
  \multicolumn{1}{c}{\textbf{(\%)}} &
  \multicolumn{1}{c}{\textbf{(\%)}} &
  \multicolumn{1}{c}{\textbf{(hours)}} &
  \multicolumn{1}{c}{\textbf{Parameter}} &
  \multicolumn{1}{c}{\textbf{gcm${-3}$}} &
  \multicolumn{1}{c}{\textbf{(n)}} &
  \multicolumn{1}{c}{\textbf{  }} \\[0.5ex]\hline \hline \\[-2ex]
\endhead
211308899&08:53:18.006&+10:01:46.97&14.57&0.56&6.423&2457143.316&0.0395&1.9868&4.885&0.151&0.45&&\\
211314705&08:36:31.182&+10:08:50.21&15.39&0.81&3.794&2457140.224&0.1007&3.1731&1.39&0&11.933&1&H$\alpha$ emission\\
211319617&08:25:51.344&+10:14:49.08&12.6&0.47&8.866&2457143.397&0.1028&3.2056&2.679&0&3.893&1&OK\\
211328748&08:40:56.495&+10:25:20.56&9.96&0.3&17.178&2457139.95&0.0219&1.4808&2.452&0&9.833&&\\
211331236&08:55:25.364&+10:28:08.87&14.65&0.86&1.292&2457140.193&0.1317&3.6291&1.153&0&7.162&&\\
211333233&08:27:35.557&+10:30:23.62&9.95&1.04&5.409&2457142.885&0.0172&1.3096&2.344&0.028&3.544&1&Giant\\
211335816&08:26:29.764&+10:33:19.44&12.2&0.4&4.99&2457140.02&0.1907&4.3667&0.862&0.487&43.754&3&SB2\\
211336616&08:49:42.413&+10:34:13.95&13.1&0.62&44.13&2457149.28&0.0427&2.0655&4.635&0.037&3.731&&\\
211342524&08:32:23.687&+10:40:38.06&12.42&0.35&14.449&2457149.367&9.1939&30.3214&0.595&0.991&1.38&&\\
211351816&08:31:03.081&+10:50:51.31&12.61&0.66&8.406&2457142.05&0.0556&2.3577&5.668&0.205&0.367&1&Giant\\
211355342&08:30:12.968&+10:54:37.06&12.75&0.41&6.893&2457143.799&0.0585&2.4185&2.468&0&3.872&1&OK\\
211359660&08:40:43.278&+10:58:58.59&12&0.49&4.737&2457141.205&0.0977&3.1257&2.41&0.042&2.852&1&OK\\
211365543&08:29:48.825&+11:05:08.36&12.12&0.25&5.264&2457143.676&0.0096&0.9804&4.185&0.148&0.59&&\\
211375488&08:43:20.828&+11:14:53.23& &1.37&2.084&2457141.721&0.2385&4.8839&1.955&0.109&2.328&&\\
211383821&08:44:09.925&+11:23:07.81&14.52&0.78&1.567&2457140.162&0.0326&1.8044&1.748&0.093&2.469&&\\
211391664&08:25:57.189&+11:30:40.12&12.17&0.26&10.136&2457145.985&0.0869&2.9486&4.945&0.171&0.678&1&OK\\
211399359&08:32:16.114&+11:37:50.62&14.64&0.58&3.115&2457141.418&2.3415&15.3019&1.992&0.049&3.322&&\\
211401787&08:27:35.269&+11:40:02.91&9.61&0.26&13.773&2457151.069&0.0245&1.5638&4.311&0.146&1.406&1&OK\\
211413752&08:54:50.291&+11:50:53.75&13.85&0.52&9.325&2457140.855&0.0923&3.0376&2.877&0&3.305&&\\
211418729&08:31:31.911&+11:55:20.15&14.56&0.53&11.391&2457140.324&1.3265&11.5172&3.384&0.108&2.438&&\\
211424769&08:35:24.644&+12:00:41.94&9.39&0.33&5.176&2457144.498&0.3485&5.9037&1.049&0.157&36.494&2&SB1\\
211428897&08:35:25.812&+12:04:33.04&14.09&0.79&1.611&2457140.661&0.0517&2.2748&1.059&0&11.471&&\\
211439059&08:47:53.647&+12:13:54.85&13.29&0.51&18.641&2457146.51&0.0285&1.6896&5.22&0.173&1.057&&\\
211442297&08:26:12.827&+12:16:54.97&13.36&0.38&20.272&2457157.159&1.4508&12.0449&3.141&0.101&5.435&&\\
211490999&08:43:11.723&+13:00:34.53&13.6&0.43&9.844&2457146.33&0.0809&2.8435&3.425&0.118&2.026&&\\
211491383&08:40:37.240&+13:00:52.83&11.78&0.33&4.144&2457141.601&0.0078&0.8845&2.717&0.101&1.722&1&OK\\
211509553&09:00:04.744&+13:16:25.94&16.58&0.8&20.359&2457151.414&3.3123&18.1997&2.938&0.131&6.602&&\\
211525389&08:21:40.866&+13:29:51.11&11.75&0.45&8.267&2457139.723&0.1053&3.2454&3.216&0.139&2.038&1&OK\\
211529065&08:45:03.983&+13:32:59.40&13.78&0.6&4.4&2457142.979&0.1297&3.6012&1.456&0&12.033&&\\
211562654&08:20:01.718&+14:01:10.06&12.85&0.46&10.792&2457147.782&0.0523&2.2862&3.581&0.151&1.916&1&OK\\
211569704&08:28:01.122&+14:07:10.62&12.36&0.72&34.023&2457156.994&0.0308&1.7559&3&0&10.625&&\\
211579112&09:03:32.001&+14:15:01.90& &0.86&17.703&2457156.427&0.6508&8.0674&1.872&0&22.768&1&OK\\
211586387&09:07:12.268&+14:21:19.71&14.64&0.65&35.383&2457142.906&0.0945&3.0738&3.094&0&10.078&&\\
211594205&08:36:33.626&+14:27:42.97&10.35&0.54&16.994&2457148.501&0.0315&1.7749&2.285&0&12.012&1&OK\\
211645912&08:19:59.301&+15:10:40.42&12.62&0.37&10.673&2457140.54&0.0274&1.6562&3.436&0.104&2.185&1&OK\\
211713099&08:20:53.731&+16:05:27.41&13.83&0.4&8.562&2457141.15&0.4581&6.768&2.975&0.109&2.696&&\\
211733267&08:40:02.259&+16:22:20.66&12.4&0.55&8.658&2457144.931&0.5656&7.5206&1.098&0.261&49.683&2&OK, $K \lesssim 100$m/s\footnote{RV measurements were 22.554 and~22.551~km/s, without uncertainties so assuming standard error.}\\
211736671&08:13:31.650&+16:25:10.59&12.33&0.43&4.734&2457140.365&0.0778&2.7891&3.407&0.197&0.956&1&Giant\\
211743874&08:37:33.528&+16:31:19.57&12.57&0.3&12.281&2457148.217&0.0233&1.5248&3.876&0.151&1.72&1&OK\\
211763214&08:55:21.136&+16:47:39.05&12.73&0.5&21.199&2457146.569&0.0182&1.3507&4.283&0.123&2.224&1&OK\\
211770696&08:31:02.684&+16:54:02.04&12.41&0.38&16.271&2457145.973&0.0312&1.7668&7.534&0.223&0.298&1&OK\\
211770795&08:48:02.336&+16:54:06.67&14.88&0.67&7.728&2457141.099&0.0791&2.8133&2.83&0&2.88&&\\
211770867&08:12:05.436&+16:54:10.82&12.28&0.3&27.693&2457147.872&1.8421&13.5722&6.343&0.465&0.636&&\\
211779390&08:17:26.659&+17:01:27.68&13.45&0.72&3.85&2457141.528&0.0282&1.6802&1.627&0&7.553&&\\
211783206&08:40:34.995&+17:04:40.57&14.6&0.65&7.134&2457146.467&0.045&2.1212&1.839&0&9.687&&\\
211800191&08:51:32.348&+17:19:11.40&12.62&0.37&1.106&2457140.749&0.0975&3.1229&0.994&0&9.572&1&OK\\
211804579&08:36:16.266&+17:22:53.98&11.36&0.39&1.523&2457141.205&0.0605&2.4602&2.397&0.156&0.92&&EB\footnote{Vanderburg, private communication.}\\
211808055&08:36:08.271&+17:25:48.30& &0.4&3.383&2457142.603&0.1641&4.0506&3.721&0.353&0.46&&\\
211814733&08:50:40.178&+17:31:29.61&11.21&0.57&14.71&2457145.892&0.4414&6.6437&2.382&0.079&9.096&&\\
211816003&08:50:29.069&+17:32:32.80&13.84&0.49&14.452&2457144.86&0.1035&3.2178&3.32&0.228&3.077&&\\
211818569&08:27:44.813&+17:34:45.83&13.32&0.69&5.186&2457143.561&1.0779&10.3821&1.778&0.133&7.584&2&OK, $K \lesssim 350$ m/s\footnote{RV uncertainties added in quadrature}. \\
211834065&08:50:11.526&+17:47:57.57&11.81&0.29&10.545&2457142.426&0.0143&1.1974&3.259&0.164&2.469&&EB\footnote{Vanderburg, private communication.} \\
211886472&09:08:31.807&+18:31:42.75&11.28&0.26&19.64&2457152.378&0.5501&7.4167&1.43&0.699&20.712&&\\
211897691&08:40:19.814&+18:41:34.51&14.66&0.62&5.75&2457142.497&0.0945&3.0749&1.255&0&24.529&&\\
211906259&08:45:14.646&+18:49:09.69&12.82&0.35&2.52&2457140.559&0.0105&1.0236&12.355&0.174&0.013&&\\
211919004&08:39:06.491&+19:00:36.08&13.37&0.51&11.72&2457149.097&0.1051&3.2422&4.406&0&1.157&&\\
211923431&08:31:44.965&+19:04:28.71&14.3&0.49&29.729&2457143.824&0.067&2.5878&4.998&0.196&1.895&&\\
211924657&08:40:06.426&+19:05:34.36&16.25&0.81&2.645&2457141.999&0.2365&4.8634&1.404&0&8.086&&\\
211929937&08:36:42.829&+19:10:25.72&14.43&0.52&3.477&2457142.412&1.7941&13.3945&2.205&0.016&2.743&&\\
211941472&08:41:47.658&+19:20:50.99&11.95&0.39&5.78&2457143.635&0.0101&1.0064&3.138&0.154&1.525&1&OK\\
211945201&09:06:17.754&+19:24:08.11&10.15&0.31&19.491&2457158.827&0.1381&3.7166&3.369&0.035&4.29&&\\
211965883&09:04:37.728&+19:42:52.51&14.63&0.82&10.555&2457146.496&0.1249&3.5337&1.307&0.3&34.611&&\\
211969807&08:38:32.821&+19:46:25.78& &0.86&1.974&2457140.376&0.1325&3.6401&1.407&0&6&&\\
211990866&08:38:24.300&+20:06:21.83&10.65&0.28&1.674&2457140.72&0.0589&2.4261&1.398&0.218&4.832&1&OK\\
211993818&08:24:49.181&+20:09:10.78&7.34&0.47&8.986&2457140.043&4.8646&22.0558&0.002&1&0.24&3&SB2\\
211995398&08:14:37.512&+20:10:44.93& & &32.576&2457169.858&3.8336&19.5795&4.253&0&3.572&&\\
212006318&08:42:00.319&+20:21:33.50&13.04&0.34&14.443&2457147.342&0.0245&1.5651&6.035&0.178&0.529&1&OK\\
212006344&08:25:54.315&+20:21:34.45&13.15&0.83&2.219&2457141.831&0.038&1.9484&1.118&0&13.438&1&Very Cool \\
212008766&08:37:07.785&+20:23:57.74&13.09&0.52&14.129&2457145.122&0.0854&2.9218&3.399&0&3.035&1&OK\\
212009427&08:31:29.870&+20:24:37.52& &0.91&0.778&2457140.262&0.5428&7.3676&0.5&0.236&48.518&&\\
212066407&08:39:21.244&+21:23:26.98&12.34&0.38&0.822&2457140.365&0.0437&2.0904&0.814&0.056&12.911&11&$K \sim 32.8 \pm 25$m/s\\
212069861&08:57:46.605&+21:27:12.72&14.78&0.85&30.953&2457147.496&0.1646&4.0566&3.236&0.152&7.442&&\\
212088059&08:50:49.887&+21:47:20.84&15.61&0.88&10.366&2457141.715&0.124&3.5211&1.892&0&12.915&&\\
212099230&08:32:17.657&+22:00:21.64&10.79&0.47&7.112&2457141.963&0.0585&2.418&2.651&0.009&3.222&1&OK\\
212110888&08:30:18.905&+22:14:09.27&11.45&0.34&2.996&2457141.351&0.6237&7.8978&1.971&0.068&3.292&5&HJ\footnote{\citet{2016arXiv160107635L} and~\citet{2016arXiv160200638H}}\\
212130773&08:23:48.660&+22:38:02.48&14.63&0.61&18.711&2457151.886&0.1305&3.6132&5.972&0.162&0.713&&\\
212132195&08:23:56.896&+22:39:46.96&12.03&0.63&26.198&2457164.39&0.0861&2.9347&3.173&0&6.919&1&OK\\
212136123&08:36:50.878&+22:44:28.04&15.09&0.5&2.226&2457140.262&0.0676&2.6003&1.679&0&3.986&&\\
212138198&08:24:05.654&+22:46:59.84&13.21&0.58&3.209&2457142.373&0.2016&4.4905&0.725&0.131&69.121&1&OK\\
212141021&08:31:41.083&+22:50:30.01&13.55&0.57&2.918&2457140.099&0.0246&1.5674&1.914&0.066&3.499&&\\
212150006&08:32:40.691&+23:01:55.20&14.77&0.51&0.898&2457139.982&0.1878&4.3334&0.958&0.122&8.524&5&\footnote{\citet{2016arXiv160306488A}}\\
212152341&08:51:00.966&+23:05:02.25& & &6.676&2457141.311&3.9424&19.8555&2.273&0.087&4.747&&\\
212154564&08:54:33.884&+23:07:58.40& &0.86&6.414&2457142.181&0.4643&6.8141&1.521&0.102&15.136&&\\
212157262&08:50:05.666&+23:11:33.36&13.08&0.49&7.15&2457146.322&0.1068&3.2683&2.803&0.03&2.739&1&OK\\
212161956&08:25:29.502&+23:17:50.56&15.21&0.75&7.187&2457140.698&0.1195&3.4576&2.391&0&4.437&&\\
212164470&08:39:15.271&+23:21:26.93&12.65&0.35&7.809&2457144.86&0.042&2.0497&3.403&0.204&1.572&1&OK\\
[0.3ex]\hline \hline \\[-2ex]
\end{longtable}
\normalsize
\renewcommand{\thefootnote}{\arabic{footnote}}
\renewcommand{\arraystretch}{1.0}
\end{center}

\begin{center}
\scriptsize
\begin{longtable}{rrrrrrrrrrrrrr}
\caption[Planet Candidates]{Full list of our 71 vetted single planet candidates in \ktwo\ C6. RV semi-amplitude $K$ (or limits) listed in Comments where applicable.}
\label{table_6} \\

\hline \hline \\[-2ex]
  \multicolumn{1}{c}{\textbf{EPIC}} &
  \multicolumn{1}{c}{\textbf{RA}} &
  \multicolumn{1}{c}{\textbf{Dec}} &
  \multicolumn{1}{c}{\textbf{V}} &
  \multicolumn{1}{c}{\textbf{J-K}} &
  \multicolumn{1}{c}{\textbf{Period}} &
  \multicolumn{1}{c}{\textbf{Epoch}} &
  \multicolumn{1}{c}{\textbf{Depth}} &
  \multicolumn{1}{c}{\textbf{$r_p/r_\star$}} &
  \multicolumn{1}{c}{\textbf{Duration}} &
  \multicolumn{1}{c}{\textbf{Impact}} &
  \multicolumn{1}{c}{\textbf{Density}} &
  \multicolumn{1}{c}{\textbf{TRES}} &
  \multicolumn{1}{c}{\textbf{Comments}}  \\[0.8ex] 
  \multicolumn{1}{c}{\textbf{  }} &
  \multicolumn{1}{c}{\textbf{  }} &
  \multicolumn{1}{c}{\textbf{  }} &
  \multicolumn{1}{c}{\textbf{(mag)}} &
  \multicolumn{1}{c}{\textbf{(mag)}} &
  \multicolumn{1}{c}{\textbf{(days)}} &
  \multicolumn{1}{c}{\textbf{(BJD)}} &
  \multicolumn{1}{c}{\textbf{(\%)}} &
  \multicolumn{1}{c}{\textbf{(\%)}} &
  \multicolumn{1}{c}{\textbf{(hours)}} &
  \multicolumn{1}{c}{\textbf{Parameter}} &
  \multicolumn{1}{c}{\textbf{gcm${-3}$}} &
  \multicolumn{1}{c}{\textbf{(n)}} &
  \multicolumn{1}{c}{\textbf{  }} \\[0.5ex]\hline \hline \\[-2ex]
\endhead
212270970&13:51:09.957&-18:27:42.22&13.48&0.42&1.717&2457218.134&0.0086&0.9286&6.192&0.185&0.065& &\\
212278644&13:48:29.897&-18:11:47.77&14.15&0.37&12.426&2457227.549&0.0431&2.077&4.923&0.213&0.821& &\\
212297394&13:48:49.493&-17:36:35.66&14.42&0.55&5.213&2457222.483&0.0709&2.6618&2.422&0.07&3.078& &\\
212300977&13:35:01.945&-17:30:12.78&11.75&0.38&4.466&2457220.529&1.5619&12.4977&3.095&0.203&1.198& &WASP-55\\
212301649&13:25:50.914&-17:28:59.27&14.33&0.56&1.225&2457217.565&0.0224&1.4962&9.016&0.554&0.019& &\\
212310244&13:52:18.702&-17:13:31.61&14.27&0.55&6.669&2457220.874&0.0213&1.4607&4.984&0.166&0.438& &\\
212321305&13:38:34.080&-16:54:36.35&14.08&0.41&34.144&2457228.926&0.2836&5.3257&7.923&0.196&0.546& &\\
212330265&13:53:31.110&-16:39:34.92&&0.9&4.174&2457220.427&0.1013&3.1828&1.863&0.019&5.453& &\\
212351026&13:26:44.850&-16:06:18.54&15.51&0.81&2.548&2457218.171&0.2483&4.9828&4.903&0.423&0.143& &\\
212351405&13:26:39.104&-16:05:42.00&14.32&0.57&2.549&2457218.158&0.1961&4.4285&3.833&0.407&0.3& &\\
212357477&13:28:03.992&-15:56:16.15&10.36&0.39&6.327&2457221.228&0.0322&1.7952&1.675&0.083&11.238&1&OK\\
212370106&13:25:05.248&-15:37:30.26&14.71&0.83&22.446&2457225.441&0.1546&3.9319&3.48&0.011&4.491& &\\
212380207&13:21:25.496&-15:22:37.99&13.37&0.44&26.147&2457241.871&0.0568&2.3842&4.341&0.094&2.66& &\\
212394689&13:34:29.110&-15:02:10.89&12.4&0.45&6.679&2457223.42&0.0655&2.5593&2.427&0.123&3.854&1&OK\\
212398508&13:34:30.927&-14:56:49.78&13.87&0.47&46.423&2457237.276&0.0681&2.6102&6.591&0.211&1.277& &\\
212418133&13:33:12.426&-14:30:14.60&13.41&0.4&3.333&2457219.846&0.0268&1.6372&3.792&0.118&0.511& &\\
212420823&13:16:23.936&-14:26:40.67&14.4&0.54&9.029&2457219.139&0.0682&2.6107&3.009&0.078&2.772& &\\
212424622&13:29:19.541&-14:21:34.07&13.35&0.33&12.012&2457219.497&0.0315&1.7745&5.162&0.153&0.712& &\\
212425103&13:37:28.700&-14:20:56.00&14.73&0.52&0.946&2457218.175&0.0301&1.7346&1.324&0.077&3.467& &\\
212432685&13:37:11.711&-14:10:50.12&13.28&0.33&1.063&2457217.992&0.0231&1.5188&1.316&0.072&3.958& &\\
212435047&13:28:31.373&-14:07:34.65&12.5&0.33&1.115&2457218.451&0.0151&1.2269&1.562&0.099&2.475&1&OK\\
212440430&13:36:08.544&-14:00:33.18&13.46&0.42&19.991&2457228.16&0.0542&2.3276&4.266&0.122&2.125& &\\
212443973&13:40:02.135&-13:55:55.34&16.03&0.79&0.779&2457217.748&0.0369&1.9207&0.66&0&23.004& &\\
212451091&14:05:53.119&-13:46:45.80&14.46&0.52&12.666&2457226.371&0.1876&4.3318&3.51&0.053&2.462& &\\
212454160&13:34:32.944&-13:42:45.42&12.88&0.42&0.876&2457218.236&0.0146&1.2103&7.372&0.526&0.032& &\\
212454422&13:14:12.755&-13:42:24.76&&0.91&3.269&2457220.227&1.2473&11.1682&0.966&0&30.571& &\\
212460519&13:34:11.169&-13:34:36.94&12.92&0.8&7.387&2457223.798&0.0858&2.9294&2.422&0.064&4.362&1&OK\\
212480208&13:41:27.239&-13:09:39.23&11.14&0.42&10.099&2457224.771&0.0179&1.3385&3.71&0.141&1.62&1&OK\\
212495601&13:15:39.034&-12:49:37.35&13.96&0.41&21.677&2457229.646&0.0605&2.4596&4.534&0.13&1.913& &\\
212496592&13:26:33.402&-12:48:23.65&13.26&0.53&2.858&2457219.561&0.0275&1.6582&2.043&0.188&2.69&1&OK\\
212499835&13:49:40.470&-12:44:17.29&16.45&0.58&6.883&2457221.015&0.4417&6.6463&5.531&0.232&0.318& &\\
212499991&13:29:57.305&-12:44:05.00&13.55&0.51&15.381&2457225.459&0.0542&2.3288&2.187&0&12.41& &\\
212521166&13:49:23.888&-12:17:04.17&11.91&0.58&13.864&2457219.875&0.1147&3.3865&3.007&0.043&4.292&1&OK\footnote{Mini-Neptune \citet{2016arXiv160504291O}.}\\
212529560&13:28:31.542&-12:06:26.34&14.09&0.3&8.121&2457219.08&0.0406&2.014&3.398&0.164&1.678& &\\
212534729&13:42:55.909&-11:59:39.30&13.29&0.49&13.479&2457223.541&0.0222&1.4904&2.84&0.01&4.966&1&OK\\
212543933&13:46:36.559&-11:48:17.80&14.15&0.37&7.806&2457223.497&0.0477&2.1838&2.75&0.042&3.159& &\\
212554013&13:48:18.812&-11:35:20.32&15.04&0.58&3.588&2457220.338&1.2649&11.2466&1.868&0.073&4.618& &\\
212555594&13:46:19.746&-11:33:22.51&12.7&0.46&4.163&2457220.44&0.0245&1.5661&1.441&0&11.741&1&OK\\
212562715&13:28:40.302&-11:23:57.59&13.28&0.42&13.524&2457220.467&0.0533&2.3085&2.314&0.191&8.706&1&OK\\
212563850&13:23:23.646&-11:22:29.11&13.71&0.77&14.311&2457222.751&0.0689&2.6248&1.96&0&16.029& &\\
212570977&13:43:36.335&-11:13:24.85&14.08&0.42&8.853&2457223.894&2.4303&15.5895&3.534&0.536&1.022& &\\
212572439&13:37:45.619&-11:11:33.26&13.18&0.53&2.581&2457217.866&0.3992&6.3185&1.51&0.052&6.319&1&HJ\footnotemark{\ref{vdb}}\\
212572452&13:37:46.022&-11:11:32.00&&0.81&2.581&2457217.865&3.029&17.404&1.509&0.153&6.14& &\\
212575828&13:40:38.328&-11:07:01.53&15.79&0.57&2.06&2457217.849&0.1344&3.6658&1.378&0&6.668& &\\
212577658&13:55:00.806&-11:04:47.35&11.8&0.45&14.069&2457221.323&0.0302&1.738&3.023&0.056&4.277&1&OK\\
212579424&13:11:25.778&-11:02:33.19&&&6.373&2457221.769&1.7919&13.3861&1.902&0&7.816& &\\
212580872&13:40:56.895&-11:00:33.47&13.25&0.46&14.786&2457224.262&0.1334&3.6518&4.117&0&1.788&1&OK\\
212586030&13:43:25.957&-10:53:48.92&12.02&0.65&7.785&2457220.882&0.0541&2.3269&1.568&0&17.048&1&OK\\
212587672&13:41:46.729&-10:51:44.75&12.49&0.32&23.226&2457237.044&0.0469&2.1656&3.038&0.399&5.382&1&OK\\
212592101&13:42:41.093&-10:46:06.70&&0.86&4.545&2457217.55&0.6354&7.9711&1.047&0.221&30.975&1&Giant\\
212639319&13:18:18.754&-09:41:30.69&12.62&0.48&13.843&2457222.438&0.0781&2.7945&1.55&0.275&27.844&1&OK\\
212645891&14:01:05.526&-09:32:24.40&12.73&0.4&0.984&2457218.042&0.1628&4.0344&0.682&0.24&24.06& &\\
212646483&14:06:39.170&-09:31:35.42&14.09&0.49&8.253&2457219.873&0.0845&2.9071&1.04&0.663&25.972&1&OK\\
212661144&13:56:56.001&-09:11:15.46&13.85&0.49&2.459&2457218.91&0.0739&2.7192&1.092&0.327&13.479&&EB\footnotemark{\ref{vdb}}\\
212672300&13:38:26.144&-08:55:37.74&12.91&0.31&39.699&2457243.008&0.0614&2.4786&7.367&0.273&0.746&1&OK\\
212679181&13:26:56.924&-08:45:46.85&12.99&0.86&1.055&2457217.551&0.0458&2.1401&0.517&0.243&58.735& &\\
212688920&13:32:03.256&-08:31:53.33&14.18&0.5&62.841&2457220.233&5.3463&23.1222&2.023&0.818&12.203&1&Cool\\
212689874&13:19:19.563&-08:30:34.13&12.5&0.37&15.854&2457225.044&0.0902&3.0041&4.502&0.112&1.439& &\\
212697709&13:26:37.247&-08:19:03.22&12.91&0.35&3.952&2457218.287&0.6205&7.8771&1.416&0.159&11.309&1&OK\\
212705192&13:30:25.305&-08:07:48.94&12.2&0.43&2.268&2457219.619&0.4633&6.8064&1.515&0.073&5.478&1&OK\\
212712473&13:56:04.647&-07:57:07.67&&1.34&7.264&2457217.793&0.3232&5.6852&3.068&0.101&2.093& &\\
212735333&13:29:34.479&-07:22:26.41&12.16&0.4&8.358&2457218.187&0.0515&2.2698&3.214&0.241&1.943&1&OK\\
212737443&13:36:53.207&-07:19:05.32&&0.66&13.603&2457221.355&0.1212&3.481&2.76&0&5.457& &\\
212756297&13:50:37.408&-06:48:14.42&13.42&0.73&1.337&2457218.11&3.1165&17.6536&1.507&0.641&1.538& &Quatar-2b\\
212757601&13:52:37.115&-06:46:09.78&&0.58&1.018&2457218.013&0.9794&9.8964&1.322&0&3.774& &\\
212772113&13:32:53.284&-06:21:24.67&14.02&0.58&8.953&2457223.806&0.1178&3.4324&0.812&0.124&137.771& &\\
212782836&13:39:07.134&-06:02:29.74&11.49&0.44&7.125&2457222.12&0.0153&1.2382&2.814&0.092&2.667& &\\
212796016&13:41:10.488&-05:39:48.11&14.85&0.81&3.216&2457217.935&0.0308&1.754&1.343&0&11.218& &\\
212803289&13:55:05.698&-05:26:32.88&11.15&0.3&18.25&2457233.823&0.1776&4.2143&10.868&0.343&0.1&3&$K \sim 50 \pm28$m/s \\
212813907&13:47:19.838&-05:06:21.87&14.3&0.56&6.725&2457221.836&0.2425&4.9246&0.921&0.156&69.908& &\\
212828909&13:29:00.385&-04:36:36.82&12.46&0.5&2.85&2457218.558&0.025&1.5815&1.33&0.132&9.965&1&OK\\
[0.3ex]\hline \hline \\[-2ex]
\end{longtable}
\normalsize
\renewcommand{\thefootnote}{\arabic{footnote}}
\renewcommand{\arraystretch}{1.0}
\end{center}

\begin{center}
\renewcommand{\thefootnote}{\fnsymbol{footnote}}
\renewcommand{\arraystretch}{0.79}
\scriptsize
\begin{longtable}{rrrrrrrrrrrrrr}
\caption[Planet Candidates]{Full list of our vetted multiple system candidates. EPIC~212012119 is in C5, while the remaining objects are all from C6.}
\label{multi_table} \\

\hline \hline \\[-2ex]
  \multicolumn{1}{c}{\textbf{EPIC}} &
  \multicolumn{1}{c}{\textbf{RA}} &
  \multicolumn{1}{c}{\textbf{Dec}} &
  \multicolumn{1}{c}{\textbf{V}} &
  \multicolumn{1}{c}{\textbf{J-K}} &
  \multicolumn{1}{c}{\textbf{Period}} &
  \multicolumn{1}{c}{\textbf{Epoch}} &
  \multicolumn{1}{c}{\textbf{Depth}} &
  \multicolumn{1}{c}{\textbf{$r_p/r_\star$}} &
  \multicolumn{1}{c}{\textbf{Duration}} &
  \multicolumn{1}{c}{\textbf{Impact}} &
  \multicolumn{1}{c}{\textbf{Density}} &
  \multicolumn{1}{c}{\textbf{TRES}} &
  \multicolumn{1}{c}{\textbf{Comments}}  \\[0.8ex] 
  \multicolumn{1}{c}{\textbf{  }} &
  \multicolumn{1}{c}{\textbf{  }} &
  \multicolumn{1}{c}{\textbf{  }} &
  \multicolumn{1}{c}{\textbf{(mag)}} &
  \multicolumn{1}{c}{\textbf{(mag)}} &
  \multicolumn{1}{c}{\textbf{(days)}} &
  \multicolumn{1}{c}{\textbf{(BJD)}} &
  \multicolumn{1}{c}{\textbf{(\%)}} &
  \multicolumn{1}{c}{\textbf{(\%)}} &
  \multicolumn{1}{c}{\textbf{(hours)}} &
  \multicolumn{1}{c}{\textbf{Parameter}} &
  \multicolumn{1}{c}{\textbf{(gcm${-3}$})} &
  \multicolumn{1}{c}{\textbf{(n)}} &
  \multicolumn{1}{c}{\textbf{  }} \\[0.5ex]\hline \hline \\[-2ex]
\endfirsthead
212012119~b&08:48:40.775&+20:27:18.27&12.063&0.615&3.281 &2457142.135 & 0.077 &2.78 &1.790 & 0 & 4.835 & 1 & \\
~~~~~~~~~~c&            &            &     &    &8.438 &2457142.495 & 0.084 &2.90&2.029 & 0.266 &7.643 & & \\
212393193~b&14:02:11.279&-15:04:10.34&11.92&0.32&14.452 &2457219.472 & 0.0331 &1.82 &2.958 & 0 & 4.711 & 1 & \\
~~~~~~~~~~c&            &            &     &    &36.152 &2457236.741 & 0.0331 &1.83 &5.317 & 0.001 &2.029 & & \\
212703473~b&13:24:56.770&-08:10:18.26&10.97&0.39&6.788& 2457222.73 & 0.0204 &1.4299 & 2.784 & 0.097 &2.622 & 1 &HJ Candidate \\
~~~~~~~~~~c&            &            &     &    &18.516 &2457221.169 & 0.03 & 1.73 & 2.62 &0.323 & 7.362 & & \\
212768333~b&13:15:22.515&-06:27:53.59&10.97&0.51& 3.359 &2457218.397 & 0.018 & 1.33 & 1.420 & 0.054 & 9.866 & 1 & Low SNR\\
~~~~~~~~~~c&            &            &     &    & 7.450 &2457221.023 & 0.0681 &2.61 &2.610 &0.003 &3.539 & & Clear\\
~~~~~~~~~~d&            &            &     &    &17.043 &2457221.61 & 0.1737 & 4.1673 & 2.581 & 0.066 & 8.305 & & Clear\\
212779596~b&13:55:36.409&-06:08:10.12&12.29&0.65&3.225  &2457218.739 & 0.058 & 2.41 & 1.827 & 0.020 & 4.469 & 1 & \\
~~~~~~~~~~c&            &            &     &    &7.374  &2457222.93 & 0.1431 & 3.7832 & 2.316 & 0.13 & 4.887 & & \\
[0.3ex]\hline \hline \\[-2ex]
\end{longtable}
\normalsize
\renewcommand{\thefootnote}{\arabic{footnote}}
\renewcommand{\arraystretch}{1.0}
\end{center}

% Don't change these lines
\bsp	% typesetting comment
\label{lastpage}
\end{document}